\documentclass[11pt]{article}
\usepackage{natbib}

\usepackage{graphicx}
\usepackage{amssymb}
\usepackage{epsfig}

\newcommand{\rem}[1]{}
\newcommand{\comment}[1]{}

\def\maps11{\stackrel {1-1}{\longmapsto}}

\newlength{\Iwidth}
\newlength{\Zwidth}

\newcommand{\qed}{%
        \hfill
	\setlength{\fboxsep}{0pt}%
	\setlength{\fboxrule}{1pt}%
	\framebox{{\rule{5pt}{6pt}}}%
}

\comment{

\newtheorem{THEOREM}{Theorem} 
\newenvironment{theorem}{\begin{THEOREM} \hspace{-.85em} {\bf :} \rm}%
                        {\end{THEOREM}} 
\newtheorem{THEOREM2}[THEOREM]{Theorem} 
                        {\end{THEOREM2}} 
\newtheorem{COROLLARY}[THEOREM]{Corollary} 
\newenvironment{corollary}{\begin{COROLLARY} \hspace{-.85em} {\bf :} \rm}%
                          {\end{COROLLARY}} 
\newtheorem{LEMMA}{Lemma} 
\newenvironment{lemma}{\begin{LEMMA} \hspace{-.85em} {\bf :} \rm}%
                      {\end{LEMMA}} 
\newtheorem{LEMMA2}[LEMMA]{Lemma}
                      {\end{LEMMA2}} 
\newtheorem{CLAIM}{Claim} 
\newenvironment{claim}{\begin{CLAIM} \hspace{-.85em} {\bf :} \rm}%
                      {\end{CLAIM}} 
\newtheorem{PROPOSITION}{Proposition} 
\newenvironment{proposition}{\begin{PROPOSITION} \hspace{-.85em} {\bf :} \rm}%
                      {\end{PROPOSITION}} 
\newtheorem{EXAMPLE}{Example} 
\newenvironment{example}{\begin{EXAMPLE} \hspace{-.85em} {\bf :} \rm}%
                        {\end{EXAMPLE}} 
\newenvironment{proof}{\noindent {\bf Proof:} \hspace{.677em}}%
                      {\qed} 
                      {} 
                      {} 
                      {}
\newtheorem{DEFINITION}{Definition} 
\newenvironment{definition}{\begin{DEFINITION} \hspace{-.85em} {\bf :} \rm}%
                           {\end{DEFINITION}} 
\newtheorem{DEFINITION2}[DEFINITION]{Definition}
                           {\end{DEFINITION2}} 
\newtheorem{OBSERVATION}{Observation} 
\newenvironment{observation}{\begin{OBSERVATION} \hspace{-.85em} {\bf :} \rm}%
                           {\end{OBSERVATION}} 

\newtheorem{REMARK}{Remark} 
\newenvironment{remark}{\begin{REMARK} \hspace{-.85em} {\bf :} \rm}%
                       {\end{REMARK}} 
\newtheorem{EXERCISE}{Exercise} 
                         {\end{EXERCISE}}

}

\usepackage{ifthen}

\newtheorem{theorem}{Theorem}
\newtheorem{definition}{Definition}

\newtheorem{proposition}{Proposition}

\newtheorem{corollary}[theorem]{Corollary}
\newtheorem{observation}{Observation}
\newtheorem{xmpl}{Example}

\newcounter{spuriouscounter}

\newtheorem{lemma*}[spuriouscounter]{Lemma}
\newtheorem{corollary*}[spuriouscounter]{Corollary}

\newenvironment{proof}{\noindent {\bfseries Proof.} }{\hfill\qed\vspace{1em}}
               {\hfill\qed\vspace{1em}}
               {\hfill\qed\vspace{1em}}
               {\hfill\qed\vspace{1em}}

\newsavebox{\saveboxtxt}

\newsavebox{\progbox}
\newsavebox{\progtitlebox}
\newlength{\titleskip}
\newlength{\codeindent}
\newlength{\linenumwidth}
\newlength{\pauseskip}
\newlength{\indentskip}

\newcounter{lynecount}
\newcounter{startflag}
\newcommand{\defaultlinenumtext}{10.\hspace{12pt}}
\newenvironment{program}[2][\defaultlinenumtext]
  {\addvspace{1em}\begin{lrbox}{\progtitlebox}{\bfseries #2}\end{lrbox}
   \setcounter{lynecount}{0}
   \setcounter{startflag}{0}
   \setlength{\pauseskip}{8pt}
   \setlength{\indentskip}{1cm}
   \setlength{\codeindent}{8mm}
   \setlength{\titleskip}{1em}
   \settowidth{\linenumwidth}{#1}
  }
  {\ifthenelse{\value{startflag} = 0}{\startprog}{}
    \end{tabbing}\end{minipage}\end{lrbox}
    \begin{center}\fbox{\usebox{\progbox}}\end{center}
    \vspace{1em}
  }

\newcommand{\startprog}
   {\setcounter{startflag}{1}
    \begin{lrbox}{\progbox}
    \begin{minipage}[b]{\textwidth}\begin{tabbing}
    \hspace{\codeindent}\=%
    \hspace{\linenumwidth}\=\hspace{\indentskip}\=%
    \hspace{\indentskip}\=\hspace{\indentskip}\=%
    \hspace{\indentskip}\=\hspace{\indentskip}\=\kill
    \usebox{\progtitlebox}\rule[-\titleskip]{0pt}{0pt}\\
   }
\newcommand{\lyne}[1][\}]
  {\ifthenelse{\value{startflag} = 0}{\startprog}{\rule{\codeindent}{0pt}\\}
   \refstepcounter{lynecount}
   \ifthenelse{\equal{#1}{\}}}{}{\label{#1}}
   \>\thelynecount .\>
  }
   {\end{minipage}\end{lrbox}\usebox{\saveboxtxt}}

\newcommand{\st}{{\hspace{1pt}:\hspace{1pt} }}

\newcommand{\sg}{\sigma}

\newcommand{\0}{\emptyset}
\newcommand{\remove}[1]{}

\newcommand{\q}{{\bf q}}
\def\hq{{\bf \hat Q}}
\def\hs{{\bf \hat S}}

\begin{document}

\author{ Benny Chor\footnote{School of Computer Science, Tel-Aviv
University, Tel-Aviv 39040 Israel. Research supported by
ISF grant 418/00. {\tt benny@cs.tau.ac.il}}
\and Michael D. Hendy\footnote{
Allan Wilson Centre for Molecular Ecology and Evolution,
Massey University, Palmerston North, New Zealand. {\tt m.hendy@massey.ac.nz}}
\and Sagi Snir\footnote{{\bf Corresponding author.}
Computer Science dept., Technion, Haifa 32000, Israel. {\tt ssagi@cs.technion.ac.il}}
}



\date{\today}

\title{ Maximum Likelihood Jukes-Cantor Triplets:\\  Analytic
Solutions}

\maketitle

\begin{abstract}
Complex systems of polynomial equations have to be set up and solved
algebraically in order to obtain analytic solutions for maximum
likelihood on phylogenetic trees. This has restricted the types of
systems previously resolved to the simplest models - three and four taxa
under a molecular clock, with just {\em two state} characters.  In
this work we give, for the first time, analytic solutions for a
 family of trees with {\em four} state characters, like normal
DNA or RNA.  The model of substitution we use is the Jukes-Cantor
model, and the trees are on three taxa under molecular clock,
namely {\em rooted triplets}.

We employ a number of approaches and tools to solve this system:
Spectral methods (Hadamard conjugation), a new representation of
variables (the {\em path-set spectrum}), and algebraic geometry tools
(the resultant of two polynomials). All these, combined with heavy
application of computer algebra packages (Maple), let us derive the
desired solution.
\end{abstract}


\section{Introduction}
\label{intro}
\begin{figure}
\begin{center}
\epsfig{file=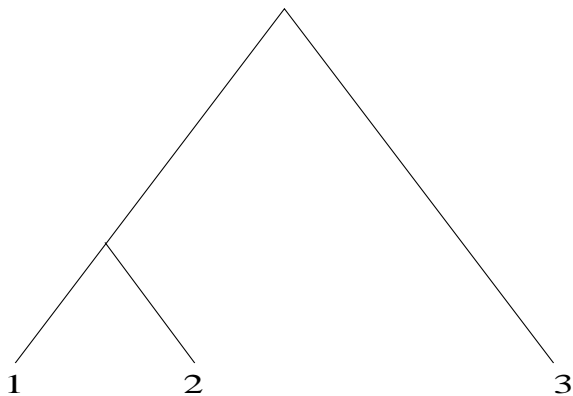,width=3in,height=5.2cm,angle=0}
\end{center}
\caption{A rooted tree over $3$ species.
\label{fig:JC-triplet}}
\end{figure}

Maximum likelihood (ML) is increasingly used as an optimality
criterion for selecting evolutionary trees \citep{Fel}, but finding the
global optimum is a hard computational task, which led to using mostly
{\em numeric} methods. So far, {\em analytic solutions} have been
derived only for the simplest models \citep{Ya00,CHP01,CKS03} -- three
and four taxa under a molecular clock, with just {\em two state}
characters \citep{Ney}.  In this work we present, for the first time,
analytic solutions for a general family of trees with {\em four} state
characters, like normal DNA or RNA.  The model of substitution we use
is the Jukes-Cantor model \citep{JC}, where all substitutions have the
same rate.  The trees we deal with are three taxa ones, namely {\em
rooted triplets} (see Figure~\ref{fig:JC-triplet}).

The change from two to four character states incurs a major increase
in the complexity of the underlying algebraic system.  Like previous
works, our starting point is to present the general maximum likelihood
problem on phylogenetic trees as a constrained optimization
problem. This gives rise to a complex system of polynomial equations,
and the goal is to solve them. The complexity of this system prevents
any manual solution, so we relied heavily on Maple, a symbolic
mathematical system. However, even with Maple, a simple attempt to
solve the system ({\em e.g.} just typing {\tt solve}) fails, and
additional tools are required.  Spectral analysis \citep{HP,HPS} uses
Hadamard conjugation to express the expected frequencies of site
patterns among sequences as a function of an evolutionary tree $T$ and
a model of sequence evolution along the edges of $T$.  As in previous
works, we used the Hadamard conjugation as a basic building block in
our method of solution. However, it turns out that the specific way we
represent the system, which is determined by the choice of variables,
plays a crucial role in the ability to solve it.  In previous works
\citep{CHP01,CKS03}, the variables in the polynomials were based on the
{\em expected sequence spectrum} \citep{HP}.  This representation leads
to a system with two polynomials of degrees $11$ and $10$. This system
is too complex to resolve with the available analytic and
computational tools. By employing a different set of variables, based
on the {\em path-set spectrum}, we were able to arrive at a simpler
system of two polynomials whose degrees are 10 and 8. To finesse the
construction, we use algebraic geometry combined with
Maple. Specifically, we now compute the resultant of the two
polynomials, which yields a single, degree $11$ polynomial in one
variable. The root(s) of this polynomial yield the desired ML
solution. We remark that similar results to the ones shown here, were
obtained by Hosten, Khetan and Sturmfels ~\citep{HKS04}, however by
using somewhat different techniques.

It is evident that the currently available heuristic methods, fail to
predict the correct tree even for small number of taxa. This is true
not only in the presence of multiple ML points, but also in cases
where the neighborhood of the (single) ML point is relatively
flat.  Therefore, a
practical consequence of this work is the use of rooted triplets in
supertree methods (constructing a big tree from small subtrees). For
unrooted trees, the subtrees must have at least four leaves ({\em
e.g. the tree from quartets} problem). For rooted trees, it is
sufficient to amalgamate a set of rooted triplets \citep{ASSU81}. Our
work enables such approaches to rely on rooted ML triplets based on
four characters states rather than two. 

Additionally,  analytic solutions are capable to 
reveal properties of the maximum likelihood points that are not
obtainable numerically. For small trees, our result can serve as a
method for checking the accuracy of the heuristic methods used in
practice.

The remainder of this work is organized as following:
In the next section we provide definitions and notations used
throughout the rest of this work.
\remove{ In Section~\ref{sec:kimura-3st} we introduce the Kimura 3ST model as a
special case of the set of substitution models named group based
models.}
In Section~\ref{sec:JC-3-seq} we develop the identities
and structures induced by the Jukes-Cantor model, while
Section~\ref{sec:ML} develops maximum likelihood on
phylogenetic trees and subsequently solves the system for the special
case of three species under
Jukes-Cantor and molecular clock. In
Section~\ref{sec:experiment} we show experimental results of applying
our method on real genomic sequences, while Section~\ref{sec:conc}
concludes with three open problems.

\newcommand{\Xb}{{\bar X}}

\section{Definitions, Notations, and the Hadamard Conjugation}
\label{sec:Hadamard}
In this section we define the model of substitution we use, introduce
useful notations, and briefly describe the Hadamard conjugation for
the Kimura models of substitutions.
\subsection{Definitions and Notations}
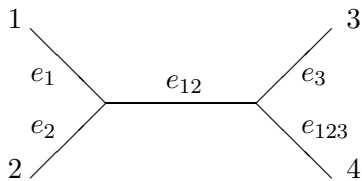
\begin{figure}[h]
\begin{center}\unitlength=1mm
\begin{picture}(55,30)
\put(18,15){\line(1,0){20}}\put(26,17){$e_{12}$}
\put(8,5){\line(1,1){10}}\put(8,11){$e_2$}
\put(8,25){\line(1,-1){10}}\put(8,18){$e_1$}
\put(38,15){\line(1,1){10}}\put(44,18){$e_3$}
\put(38,15){\line(1,-1){10}}\put(44,11){$e_{123}$}
\put(5,25){1}
\put(5,5){2}
\put(50,25){3}
\put(50,5){4}
\end{picture}\\
\end{center}
\caption{The tree $T=(12)(34)$ and its edges}
\label{fig:quartet}
\end{figure}

We start with a tree labelling notation that will be useful  for
the rest of the work. We illustrate it for $n=4$ taxa, but the
definitions extend to any $n$. A {\em split} of the species is any
partition of $\{1,2,3,4\}$ into two disjoint subsets. We will
identify each split by the subset which does {\em not} contain $4$
(in general, $n$), so that for example the split
$\{\{1,2\},\{3,4\}\}$ is identified by the subset $\{1,2\}$. For
brevity, to label objects in a split, we concatenate the members
of the split.
 Each edge $e$ of a phylogenetic tree $T$ induces a split
of the taxa, which is the cut induced by removing $e$. We denote the edge
$e$ by the cut it induces.
For instance the central edge of the tree $T\,=\,(12)(34)$
induces the split $\{\{1,2\},\{3,4\}\}$, that is
identified by the subset $\{1,2\}$ and therefore
this edge is denoted $e_{12}$. Thus the set of edges of T is
$E(T)=\{e_1,e_2,e_{12},e_3,e_{123}\}$ (see Figure~\ref{fig:quartet}).

Throughout the paper, we will index our vectors and matrices by a
method denoted {\em subsets indexing}.  We encode a subset of
$\{1,2,\ldots,n\}$ in an $(n)$-long binary number where the $i$th
least significant bit ($i=1,\ldots,n$) is "1" if $i$ is in the subset,
and "0" otherwise. Using this representation, it is convenient to
index the rows and columns of a matrix by subsets of
$\{1,2,\ldots,n\}$ in a lexicographically increasing order
(i.e. $\phi,\{1\},\{2\},\{1,2\},\ldots,\{1,2,\ldots,n\}$).
Table~\ref{tab:subsets-indexing} illustrates a matrix $M$ indexed by
subsets indexing over the set $\{1,2\}$. The general element of $M$
corresponding to subsets $D$ and $E$, is
denoted be $m_{D,E}$.

\begin{table}[hb]
\begin{center}
{\small
\begin{tabular}{|r||c|c|c|c|}
\hline
 &&&&\\
 &\{\} & \{1\}& \{2\}& \{1,2\}\\
 & 00 & 01 & 10 & 11 \\
\hline
\hline
\{\} 00 & &  &  &  \\
\hline
\{1\} 01 & &&&\\
\hline
\{2\} 10 & &&&$m_{2,12}$\\
\hline
\{1,2\} 11 & &&&\\
\hline
\end{tabular}
} 
\end{center}
\caption{The matrix M indexed by split indexing. The element
$m_{\{2\},\{12\}}$ is placed in the $(2,3)$ (binary $(10,11)$) entry.}
\label{tab:subsets-indexing}
\end{table}

Extending the alphabet from two to four character states significantly
increases the complexity of handling the data.
In contrast to the binary case, as treated in previous works
\citep{Ya00,CHP01,CKS03,CS04}, where each site pattern in the sequence data induced a split, the four
state site patterns induce a pair of splits. We will use the term
{\em substitution pattern} to represent the substitutions to each taxon from a
reference taxon. Let $X=\{1,2,\cdots,n\}$ represent the set of
taxa under study. We select $n$ as a reference
taxon and let
$\Xb=X-\{n\}$ the set of non-referenced taxa.  Consider a $n$-dimensional vector $\nu$ over the
DNA alphabet, where each entry $i$ correspond to a taxon $i$ in
$X$. The vector $\nu$ is called a character pattern. A substitution
pattern is a ($n-1$)-dimensional vector of the substitution types
$\nu_{n} \rightsquigarrow \nu_{i}$ for $i\in \Xb$.  \\
For example, the character pattern $\left [
  \begin{array}{c}A\\C\\T\\T\end{array} \right ] $ induces the
substitution pattern $\left [ \begin{array}{c} T \rightsquigarrow A \\
    T \rightsquigarrow C \\ T \rightsquigarrow T \\ \\ \end{array} \right ] $.

\remove{

We select $n$ as a reference
taxon and let
$X^*=X-\{n\}$ the non-referenced taxa.
$\nu$ and the state at the reference taxon, $\nu_n$, define a $n-1$
dimensional vector of substitutions: $\nu_{n} \leftrightarrow
\nu_{i}$ for $i= 1\ldots n-1$.

}

Suppose a phylogenetic tree $T$ over the set of taxa $X$ is given, with
substitution probabilities on each of its edges. Then, the
probability of obtaining each substitution pattern is well defined. We
remark that the number of substitution pattern is $\Sigma \times
\Sigma^{n-1}= \Sigma^n$. For some popular models, the set
of substitutions is substantially smaller than the general case.

\begin{figure}\unitlength=1.5mm
\begin{center}
\begin{picture}(60,42)
\put(10,30){\makebox(0,0){(a)}} \put(40,30){\makebox(0,0){(b)}}
\put(0,0){\makebox(0,0){\tt A}} \put(20,0){\makebox(0,0){\tt G}}
 \put(0,20){\makebox(0,0){\tt U(T)}} \put(20,20){\makebox(0,0){\tt C}}
\put(4,1){\vector(1,0){12}} \put(16,-0.5){\vector(-1,0){12}}
\put(4,19){\vector(1,0){12}} \put(16,20.5){\vector(-1,0){12}}
\put(1,4){\vector(0,1){12}} \put(-0.5,16){\vector(0,-1){12}}
\put(19,4){\vector(0,1){12}} \put(20.5,16){\vector(0,-1){12}}
\put(4.5,3.5){\vector(1,1){12.5}} \put(16,17){\vector(-1,-1){12.5}}
\put(16,3.5){\vector(-1,1){12.5}} \put(4.5,17){\vector(1,-1){12.5}}
\put(10,-2){\makebox(0,0){$\alpha$}} \put(10,22){\makebox(0,0){$\alpha$}}
\put(-2,10){\makebox(0,0){$\beta$}} \put(22,10){\makebox(0,0){$\beta$}}
\put(6,10){\makebox(0,0){$\gamma$}} \put(14,10){\makebox(0,0){$\gamma$}}

\put(30,0){\makebox(0,0){$(1,0)$}} \put(50,0){\makebox(0,0){$(1,1)$}}
 \put(30,20){\makebox(0,0){$(0,0)$}} \put(50,20){\makebox(0,0){$(0,1)$}}
\put(34,1){\vector(1,0){12}} \put(46,-0.5){\vector(-1,0){12}}
\put(34,19){\vector(1,0){12}} \put(46,20.5){\vector(-1,0){12}}
\put(31,4){\vector(0,1){12}} \put(29.5,16){\vector(0,-1){12}}
\put(49,4){\vector(0,1){12}} \put(50.5,16){\vector(0,-1){12}}
\put(34.5,3.5){\vector(1,1){12.5}} \put(46,17){\vector(-1,-1){12.5}}
\put(46,3.5){\vector(-1,1){12.5}} \put(34.5,17){\vector(1,-1){12.5}}
\put(40,-2){\makebox(0,0){\small $t_\alpha$}} \put(40,22){\makebox(0,0){\small  $t_\alpha$}}
\put(27,10){\makebox(0,0){\small $t_\beta$}} \put(53,10){\makebox(0,0){\small $t_\beta$}}
\put(35,10){\makebox(0,0){\small $t_\gamma$}} \put(45,10){\makebox(0,0){\small $t_\gamma$}}

 \end{picture}
\end{center}

\caption{ (a) Kimura's 3--substitution model (K3ST).
(b) Substitution types $t_\alpha=t_{01}$, $t_\beta=t_{10}$,
$t_\gamma=t_{11}$ and $t_\epsilon=t_{00}$.
}\label{fig:kimura}
\end{figure}
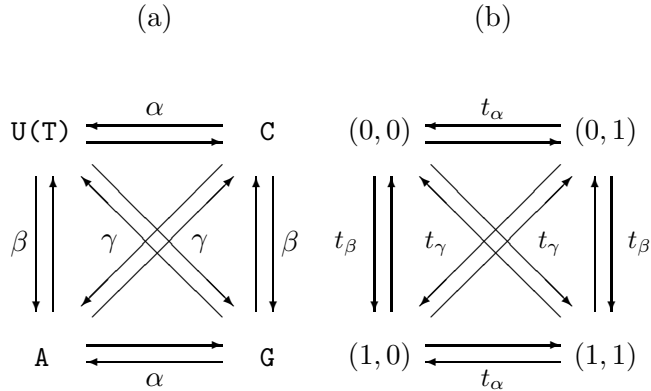

The Kimura 3 substitution model~\citep{Kimura1983}, is a model of symmetric nucleotide
substitutions, implying convergence to equal base frequencies.
In that model, Kimura proposed 3 classes of substitution: transitions
(denoted $\alpha$, $A\leftrightsquigarrow G$, $T\leftrightsquigarrow
C$), type I transversions (denoted $\beta$, $A\leftrightsquigarrow T$,
$G\leftrightsquigarrow C$) and type II transversions (denoted
$\gamma$, $A\leftrightsquigarrow C$, $T\leftrightsquigarrow G$).
Figure~\ref{fig:kimura} illustrates these relations.  We denote each
of the substitution types with a pair of binary numbers:
$t_\alpha=t_{01}$ for transitions, $t_\beta=t_{10}$, $t_\gamma=t_{11}$
for transversions and we write $t_\varepsilon=t_{00}$ for no substitution.

The number of substitution patterns with this coding is $4^{n-1}$
(for every taxon, the substitution $\nu_n \rightarrow \nu_i$ is either
of type $t_\alpha$, $t_\beta$, $t_\gamma$ or $t_\varepsilon$).

 We now define two subsets, $D,E \subseteq \Xb$,
as follows: $D=\{i \st \nu_{n} \rightarrow \nu_{i}\in \{t_\beta, t_\gamma\}\}$ and $E=\{i
\st \nu_{n} \rightarrow \nu_{i}\in \{t_\alpha, t_\gamma\}\}$. Since
both $D$ and $E$ contain species with substitution type $t_\gamma$,
they are {\em not} disjoint. To better
understand this classification of the species into the sets $D$ and
$E$, we define an encoding   of the  character states as
follows:
\begin{eqnarray*}
A&\rightarrow&(1,0)\\
C&\rightarrow&(0,1)\\
G&\rightarrow&(1,1)\\
T&\rightarrow&(0,0)
\end{eqnarray*}

With this mapping, $D$
contains the
species $i$ such that the first bit that encodes the state of $i$
differs from the first bit that encodes the state of the reference
species, $n$. The set $E$
 contains all
species $i$ such that the second bit that encodes the state of $i$
differs from the second bit that encodes the state of species
$n$. For example, suppose the character pattern $\nu$ is as follows:\\\\

\begin{tabular}{cccccc}
species ($i$) & state ($\nu_{i}$) &  binary encoding &substitution &
 membership in $D$ &
 membership in  $E$\\
--------------&--------------&---------------------&-----------------&-----------------------&--------------------------\\
1 & A                  & (1,0)     &$t_\beta=(1,0)$           &     1  &   0\\
2 & C                  & (0,1)    &$t_\alpha=(0,1)$          &   0     &  1  \\
3 & G                  & (1,1)    &$t_\gamma=(1,1)$         &  1  & 1   \\
4 & T                   & (0,0)   &        &   & \\
\end{tabular}

We can view the set $D$ (resp. $E$) as a split $\{D,\Xb\setminus D\}$
(resp. $\{E,\Xb\setminus E\}$).  We encode every substitutions pattern
by the two ordered splits $(D,E)$ that define it.  Let $s_{D,E}$ be
the probability of obtaining the substitution pattern $(D,E)$ on a
tree.  Both $D$ and $E$ range over all subsets of $\Xb$. Therefore it
is natural to represent all probabilities $s_{D,E}$ in a matrix $S=
[s_{D,E}]$, indexed by subsets indexing over $\Xb\times \Xb$. The rows
are indexed by the split $D$ and the columns
by the split $E$. We call the matrix $S$ the {\em expected sequence
spectrum}. Since the number of splits over $\Xb$ is $2^{n-1}$, $S$ is
a $2^{n-1}\times 2^{n-1}$ matrix.

For an edge $e$, let $q_e(\alpha)$, $q_e(\beta)$ and $q_e(\gamma)$ be
the expected number of substitutions of type $\alpha$, $\beta$, and
$\gamma$, respectively. We call them the {\em edge length} parameters,
so each edge is associated with three different ``lengths'', one per
substitution type. Tree edges naturally correspond to splits. We extend the
notion of edge lengths to splits that do not correspond to tree
edges, by simply defining the length as zero: For a subset $D
\subseteq \Xb$ such that $D\ne\0$ and $D$ is not an
edge split, we set $q_{D}(\theta)=0, (\theta \in
\{\alpha,\beta,\gamma\})$. For $D=\emptyset$, we set
$q_{\0}(\theta)=-K(\theta)$ where $K(\theta)$ is the sum of all other
$q_{D}(\theta)$ values. We define three vectors ${\bf q}_\theta$ for
$\theta=\alpha,\beta,\gamma$ indexed by subsets indexing over $\Xb$ as
follows: ${\bf q}_\theta=[q_D(\theta)|D \subseteq \bar X]$ . Then
$q_D(\theta)=0$ implies there is no edge $e_D$ in $T$ (e.g. ${\bf
q}_{13}(\theta)$, ${\bf q}_{23}(\theta)$ in
$T_{12}$). Figure~\ref{hadamard-fig3}(a) shows the edge length spectra
for the tree $T_{12}$ on $n=4$ taxa that was illustrated in Figure
\ref{fig:quartet}.

We will find it convenient to put these three vectors into a matrix $Q
(=Q_T)=[q_{D,E}]$ of $2^{n-1}$ rows and columns indexed by subsets
indexing over $\Xb\times \Xb$, with $q_{\0,\0}=-(K(\alpha)+K(\beta)+K(\gamma))$,
and the remaining entries of ${\bf q}_\alpha$, ${\bf q}_\beta$ and
${\bf q}_\gamma$ becoming the leading row, column and main diagonal of
$Q$ respectively. All other entries of $Q$ are set to $0$.
Figure~\ref{hadamard-fig3}(b) shows the matrix $Q=Q_{T_{12}}$ holding
the vectors ${\bf q}_\alpha$, ${\bf q}_\beta$, ${\bf q}_\gamma$ from
Figure~\ref{hadamard-fig3}(a).
\remove {These
vectors are placed in the leading row, column and main diagonal of the
matrix $Q$.
} 
This means that for $D,E \subseteq \{1,2,3\}$,
$Q_{\0,D}=q_D(\alpha)$, $Q_{D,\0}=q_D(\beta)$, $Q_{D,D}=q_D(\gamma)$,
and for all other entries, $Q_{D,E}=0$, except the
first entry $Q_{\0,\0}=-(K(\alpha)+K(\beta)+K(\gamma))$.
The entries indicated by ``$\cdot$" are all zero, and are zero for every tree.
The entries indicated by ``$0$" are zero for this specific tree
$T_{12}$, but for
different trees can be non-zero.
The non-zero entries (in the leading row, column and main diagonal)
should each be in the same component, and these identify the edge
splits of $T$.
For general trees on $n$ taxa, the edge length spectra are vectors
and square matrices of order $2^{n-1}$.

\begin{figure}
\begin{center}
\begin{tabular}{c}
$
{\bf q}_\alpha=\left[\begin{array}{r}
-K(\alpha)\\q_{1}(\alpha)\\q_{2}(\alpha)\\q_{12}(\alpha)\\q_{3}(\alpha)\\0
\\0\\q_{123}(\alpha)\end{array}\right],
{\bf q}_\beta=\left[\begin{array}{r}
-K(\beta)\\q_{1}(\beta)\\q_{2}(\beta)\\q_{12}(\beta)\\q_{3}(\beta)\\0
\\0\\q_{123}(\beta)\end{array}\right],
{\bf q}_\gamma=\left[\begin{array}{r}
-K(\gamma)\\q_{1}(\gamma)\\q_{2}(\gamma)\\q_{12}(\gamma)\\q_{3}(\gamma)\\0
\\0\\q_{123}(\gamma)\end{array}\right],$
\\\\ (a) \\ \\\\
$Q_T=\left[\begin{array}{rccccccc}
-K&q_{1}(\alpha)&q_{2}(\alpha)&q_{12}(\alpha)&q_{3}(\alpha)&0&0&q_{123}(\alpha)\\
q_{1}(\beta)&q_{1}(\gamma)&.&.&.&.&.&.\\
q_{2}(\beta)&.&q_{2}(\gamma)&.&.&.&.&.\\
q_{12}(\beta)&.&.&q_{12}(\gamma)&.&.&.&.\\
q_{3}(\beta)&.&.&.&q_{3}(\gamma)&.&.&.\\
0&.&.&.&.&0&.&.\\
0&.&.&.&.&.&0&.\\
q_{123}(\beta)&.&.&.&.&.&.&q_{123}(\gamma)
\end{array}\right],$ \\\\
(b)
\end{tabular}
\caption{(a): Example edge length spectra for the
tree $T_{12}$.  (b): $Q=Q_{T_{12}}$ }
\label{hadamard-fig3}
\end{center}
\end{figure}

\subsection{Hadamard Conjugation}

The Hadamard conjugation \citep{HP,HPS} is an invertible transformation
that specifies a relation between the expected
sequence spectrum $S$ and the edge lengths  spectra ${\bf q}(\theta)$
of the tree.
In other words, the transformation
links the probabilities of site substitutions on edges of an
evolutionary tree $T$  to the probabilities of
obtaining each possible substitutions pattern.
The Hadamard conjugation is applicable to a number of site
substitution models:  Neyman 2 state model, Jukes--Cantor
model \citep{JC}, and Kimura 2ST and 3ST models \citep{Kimura1983} (the
last three models correspond to  four states characters, such as
DNA or RNA). For these models, the
transformation yields a powerful tool which greatly simplifies and
unifies the analysis of phylogenetic data, and in particular the
analytical approach to ML.
\begin{definition}
A {\em Hadamard matrix} of order $\ell$ is an $\ell\times \ell$ matrix
$A$ with
$\pm 1$ entries such that $A^tA=\ell I_\ell$.
\end{definition}
We will use a special family of Hadamard matrices,
called Sylvester matrices in MacWilliams and Sloan
(1977, p. 45), defined inductively for
$n\geq 0$ by $H_0=[1]$ and
$H_{n+1}=\left[ \begin{array}{rr}
                         H_n & H_n\\
                         H_n & -H_n
                 \end{array}\right] . $
For example,
$$H_1=\left[ \begin{array}{rr}
                         1 &  1 \\
                         1 & -1\\
             \end{array}\right]
  \mbox{ and }
H_2=\left[ \begin{array}{rrrr}
                         1 &  1 & 1 &  1\\
                         1 & -1 & 1 & -1\\
                         1 &  1 &-1 & -1\\
                         1 & -1 &-1 &  1
                 \end{array}\right]   . $$

$H_n$ is indexed by subsets indexing over
$\{1,\ldots,n\}\times\{1,\ldots,n\}$. Let $h_{D,E }$ be the general
element of $H_n$. Then:

\begin{observation}
 $h_{D,E}=(-1)^{|D \cap E|}$.
\end{observation}

 This implies
that $H_{n}$ is symmetric, namely $H_{n}^t=H_{n}$, and thus by
the definition of Hadamard matrices $H_{n}^{-1}=\frac{1}{2^{n}}H_{n}$.

\begin{proposition}(Hendy and Penny  1993)\label{Hadam}
\space  Let $T$ be a phylogenetic tree on $n$ leaves with finite edge
lengths ($q_e(\theta)< \infty$ for all $e\in E(T)$ and $\theta \in
\{\alpha,\beta,\gamma\}$).
Assume that sites mutate according to a symmetric substitution model,
with equal
rates across sites. Let $S$ be the expected sequence spectrum and $Q$
the edge length spectrum as was described above.  Then
\begin{equation}\label{eq:had}
S = S(Q) = H_{n-1}^{-1}\exp(H_{n-1}Q)\ ,
\end{equation}

where the exponentiation function $exp(x)=e^x$ is applied element wise
to  the   matrix  $R = H_{n-1}Q$.
\end{proposition}
This transformation is called the {\em  Hadamard conjugation}.\\
\begin{definition}
 A matrix  ${\hs}\in {\mathbb R}^{2^{n-1}}\times{\mathbb R}^{2^{n-1}}$
satisfying
$\sum_{D,E\subseteq\{1,\ldots ,n-1\}} \hs_{D,E} =1$
 and $H_{n-1}\hs> {\bf 0}$ is called  {\em conservative}.
\end{definition}
 For conservative data $\hs$, the
 Hadamard conjugation  is invertible, yielding : $$\hq = \hq(\hs) =
 H_{n-1}^{-1}\ln(H_{n-1}\hs)\ $$
where the ln function is applied element-wise to  the
matrix $ H_{n-1}\hs$. We note that $\hq$ is not necessarily the edge
length  spectrum of any tree. On the other hand,
the expected sequence spectrum of any tree $T$ is always conservative.

Consider now a set of $n$ aligned homologous sequences $ \sg_1,
\cdots, \sg_n$, and denote this alignment as $AL$ . We can view $AL$
as a table where each column in this table induces a substitution
pattern. Let $F_{D,E}$ be the frequency of the substitution pattern
represented by the splits $(D,E)$. The matrix $F=[F_{D,E}]$ is denoted
as the {\em observed sequence spectrum} and is indexed analogously to the
expected sequence spectrum matrix, $S$ (that is, by subset indexing
over $\Xb\times \Xb$).

Table~\ref{tab:table1}(a) illustrates four sample DNA sequences
with sixteen sites. $\sg_4$ is the {\em reference} sequence, the
pair of binary digits above each character of $\sg_1, \cdots,
\sg_3$ is the substitution type to derive that character from the
homologous character of $\sg_4$. For example, the entry {\tt 11}
above {\tt G} at site $10$ of $\sg_1$ indicates that the
substitution to this nucleotide from the corresponding {\tt T} of
the reference sequence $\sg_4$ is of type $t_\gamma$. In (b), the
frequencies of each of the site patterns from (a) are summarized
in the observed sequence spectrum $F$. The rows of $F$ are indexed
by the first triple of the binary pairs, and the columns by the
second, in the order $000, 001, 010, 011, 100, 101, 110, 111.$ The
site pattern of site $10$ is represented by the pair $(101,001)$
(or $D=\{1,3\}$, $E=\{1\}$ alternatively) so the entry
corresponding to this is in row $101$ and column $001$ of $F$. As
this pattern occurs only at site $10$, the entry in row $101$ and
column $001$ of $F$ is {\bf 1} (highlighted in {\bf bold font}).
We emphasize that the examples here refer to a tree on four
leaves. The trees we solve for in the next sections have only {\em
three} leaves.

\begin{table}
\begin{tabular}{c}
$\begin{array}{c}
\small{
\begin{tabular}{rrrrrrrrrrrrrrrrr}
site &1&2&3&4&5&6&7&8&9&10&11&12&13&14&15&16\\
\hline
&{\tt 11}&{\tt 00}&{\tt 00}&{\tt 11}&{\tt 01}&{\tt 00}&{\tt 01}&{\tt 10}&{\tt 10}&{\tt 11}&{\tt 00}&{\tt 01}&{\tt 01}&{\tt 10}&{\tt 00}&{\tt 10}\\
$\sg_1=$& {\tt C }&{\tt C }&{\tt A}&{\tt T}&     {\tt C}&{\tt A}&    {\tt A}&{\tt A}&    {\tt C}&{\tt G}
& {\tt T }& {\tt G }& {\tt T }& {\tt G }& {\tt A }& {\tt C }\\
               &{\tt 00}&{\tt 00}&{\tt 00}&{\tt 00}&{\tt 01}&{\tt 00}&{\tt 01}&{\tt 00}&{\tt 00}&{\tt 00}
&{\tt 00}&{\tt 00}&{\tt 01}&{\tt 00}&{\tt 01}&{\tt 10}\\
$\sg_2=$& {\tt A }&{\tt C }&{\tt A}&{\tt G}&     {\tt C}&{\tt A}&   {\tt A}&{\tt T}&    {\tt G}&{\tt T}
& {\tt T }& {\tt A }& {\tt T }& {\tt C }& {\tt T }& {\tt C}\\
               &{\tt 11}&{\tt 00}&{\tt 00}&{\tt 11}&{\tt 00}&{\tt 01}&{\tt 01}&{\tt 10}&{\tt 00}&{\tt 10}
&{\tt 00}&{\tt 01}&{\tt 00}&{\tt 10}&{\tt 01}&{\tt 11}\\
$\sg_3=$& {\tt C }&{\tt C }&{\tt A}&{\tt T}&     {\tt T}&{\tt G}&    {\tt A}&{\tt A}&    {\tt G}&{\tt A}
& {\tt T }& {\tt G }& {\tt C }& {\tt G }& {\tt T }& {\tt T }\\\\
$\sg_4=$& {\tt A }&{\tt C }& {\tt A}&{\tt G}&    {\tt T}&{\tt A}&    {\tt G}&{\tt T}&    {\tt G}&{\tt T}
& {\tt T }& {\tt A }& {\tt C }& {\tt C}& {\tt A }& {\tt G }\\
\hline
\end{tabular}}\\
a
\end{array} $
\\\\
$\begin{array}{c}
\small{
F=\left[\begin{array}{rrrrrrrr}
3&0&0&2&1&1&1&1\\
1&0&0&0&0&0&0&0\\
0&0&0&0&0&0&0&0\\
0&0&0&0&0&0&0&0\\
0&0&0&0&0&0&0&0\\
2&{\mathbf 1}&0&0&0&2&0&0\\

0&0&0&0&0&0&0&0\\
0&0&0&0&1&0&0&0
 \end{array}\right]
}
\\
{ \large {b}}
\end{array}
$
\end{tabular}
 \caption{(a):Four aligned sequences with sixteen sites. (b): The
 corresponding observed sequence spectrum}
\label{tab:table1}
\end{table}

\section{Jukes--Cantor model for 3 sequences}
\label{sec:JC-3-seq}
The Jukes--Cantor model of evolution \citep{JC} is the simplest model
for four states DNA evolution. The assumption in this model is that
when a base
changes, it has equal probabilities to change to each of the other
three bases. This model can be derived from the more general Kimura
$3-$ST model by setting, for each edge of $T$, each of the three edge
length parameters equal to a common value, namely setting
$q_e(\alpha)=q_e(\beta)=q_e(\gamma)=q_e$. We now look on the tree $T$
on three taxa $\{0,1,2\}$ before determining where the root is.  $T$
has just one topology, the star with the three edges $e_1$, $e_2$ and
$e_{12}$.  For convenience we will write the edge length of $e_{12}$
as $q_3$.

We now define several auxiliary matrices that will be useful in the sequel:

$$H=\left[\begin{array}{rrrr}1&1&1&1\\1&-1&1&-1\\1&1&-1&-1\\1&-1&-1&1\end{array}\right],
J=\left[\begin{array}{cccc}1&1&1&1\\1&1&1&1\\1&1&1&1\\1&1&1&1\end{array}\right],
A_0=\left[\begin{array}{cccc}1&0&0&0\\0&0&0&0\\0&0&0&0\\0&0&0&0\end{array}\right],
A_1=\left[\begin{array}{cccc}0&1&0&0\\1&1&0&0\\0&0&0&0\\0&0&0&0\end{array}\right],$$
$$A_2=\left[\begin{array}{cccc}0&0&1&0\\0&0&0&0\\1&0&1&0\\0&0&0&0\end{array}\right],
A_3=\left[\begin{array}{cccc}0&0&0&1\\0&0&0&0\\0&0&0&0\\1&0&0&1\end{array}\right],
L=J-A_0-A_1-A_2-A_3=
\left[\begin{array}{cccc}0&0&0&0\\0&0&1&1\\0&1&0&1\\0&1&1&0\end{array}\right].
$$
The following identities relating these seven matrices, hold:
\begin{equation}\label{eq1}HJH=16A_0,\end{equation}
\begin{equation}\label{eq2}HA_0H=J,\end{equation}
\begin{equation}\label{eq3}HA_1H=4(A_0+A_2)-J,\end{equation}
\begin{equation}\label{eq4}HA_2H=4(A_0+A_1)-J,\end{equation}
\begin{equation}\label{eq5}HA_3H=4(A_0+A_3)-J.\end{equation}

The edge--length spectrum of an arbitrary 3-tree can be expressed as
the $4 \times 4$ matrix,
\begin{eqnarray*}
Q&=&\left[\begin{array}{cccc}
-3(q_1+q_2+q_3)&q_1&q_2&q_3\\q_1&q_1&0&0\\q_2&0&q_2&0\\q_3&0&0&q_3\end{array}\right]
=q_1A_1+q_2A_2+q_3A_3-3(q_1+q_2+q_3)A_0.
\end{eqnarray*}
Now, from Equations \ref{eq1}--\ref{eq5} we see
\begin{eqnarray*}
HQH&=&-4[(q_1+q_3)A_1+(q_2+q_3)A_2+(q_1+q_2)A_3+(q_1+q_2+q_3)L]\\\\
&=&-4\left[\begin{array}{cccc}
0&q_1+q_3&q_2+q_3&q_1+q_2\\
q_1+q_3&q_1+q_3&q_1+q_2+q_3&q_1+q_2+q_3\\
q_2+q_3&q_1+q_2+q_3&q_2+q_3&q_1+q_2+q_3\\
q_1+q_2&q_1+q_2+q_3&q_1+q_2+q_3&q_1+q_2\end{array}\right],
\end{eqnarray*}
so applying the exponential function to each element of the matrix
$HQH$ we obtain the so called path--set spectrum, $R$:
\begin{eqnarray}
\label{eq:path-set}
R&=&\exp(HQH)\nonumber\\
&=&A_0+x_1x_3A_1+x_2x_3A_2+x_1x_2A_3+x_1x_2x_3L\nonumber\\
&=&\left[\begin{array}{cccc}
1&x_1x_3&x_2x_3&x_1x_2\\x_1x_3&x_1x_3&x_1x_2x_3&x_1x_2x_3\\
x_2x_3&x_1x_2x_3&x_2x_3&x_1x_2x_3\\x_1x_2&x_1x_2x_3&x_1x_2x_3&x_1x_2\end{array}\right],
\end{eqnarray}
where \begin{equation} x_i = e^{-4q_i}.\end{equation}
The $x_i$ values can replace the $q_i$ values as the defining
parameters and are called the {\em path set variables}. The entries of $R$
relate to the probabilities of differences between the end-points of paths
in $T$.

By  using Proposition~\ref{Hadam}, the expected sequence spectrum equals
\begin{eqnarray}
S&=&H^{-1}RH^{-1}\\
&=&\frac1{16}\left [(1+3x_1x_2+3x_1x_3+3x_2x_3+6x_1x_2x_3)A_0\nonumber
\right .\\
 &&\left . +(1-x_1x_2-x_1x_3+3x_2x_3-2x_1x_2x_3)A_1\nonumber \right .\\
 &&\left .+(1-x_1x_2+3x_1x_3-x_2x_3-2x_1x_2x_3)A_2\nonumber \right .\\
 && \left .+(1+3x_1x_2-x_1x_3-x_2x_3-2x_1x_2x_3)A_3\nonumber\right . \\
 &&\left .+(1-x_1x_2-x_1x_3-x_2x_3+2x_1x_2x_3)L\right]\nonumber \\
&=&\frac1{16}\left[\begin{array}{cccc }
a_0&a_1&a_2&a_3\\a_1&a_1&a_4&a_4\\a_2&a_4&a_2&a_4\\a_3&a_4&a_4&a_3 \label{eq16a}
\end{array}\right],\end{eqnarray}
where 
\begin{eqnarray} a_0&=&(1+3x_1x_2+3x_1x_3+3x_2x_3+6x_1x_2x_3),\nonumber \\
a_1&=&(1-x_1x_2-x_1x_3+3x_2x_3-2x_1x_2x_3),\nonumber \\
a_2&=&(1-x_1x_2+3x_1x_3-x_2x_3-2x_1x_2x_3),\nonumber \\
a_3&=&(1+3x_1x_2-x_1x_3-x_2x_3-2x_1x_2x_3),\nonumber \\
a_4&=&(1-x_1x_2-x_1x_3-x_2x_3+2x_1x_2x_3).\label{eq16b}
\end{eqnarray}Thus we see that each expected sequence frequency takes one of the above values, which
are functions of the three parameters $x_1$, $x_2$ and $x_3$.

\section{Obtaining the Maximum Likelihood Solution}
\label{sec:ML}
\remove{
Given the observed the frequencies, $F_{DE}$, of each site pattern $(D,E) \subseteq \Xhbh\times \Xb$
(normalised so that $\sum_{D,E \subseteq \Xb}F_{D,E}=1$), then for any
expected sequence spectrum $S$ of some tree $T$,
the likelihood of obtaining those normalised frequencies is
\begin{equation}
\label{eq15}
L(F|T,S)=\prod_{D,E \subseteq \Xb}S_{D,E}^{F_{D,E}}.
\end{equation}
(It is convenient to use normalised frequencies as it simplifies the
formulae later.  This normalisation scales $\log L$ by a constant
factor, so it does not affect the identity of the turning points.)
The expected sequence spectrum $S$ is a function of the $3$ variables
$x_1$, $x_2$ and $x_3$, so the values which maximise the likelihood
$L$ will occur when the three partial derivatives are zero,
$\frac{\partial L}{\partial x_i}=0.$ By the chain rule
$$\frac{\partial L} {\partial{x_i}}=L\cdot\sum_{D,E \subseteq
\Xb}\frac{F_{D,E}}{S_{D,E}}\frac{\partial
S_{D,E}}{\partial{x_i}}\hspace{1cm} (i=1,2,3),$$ so the turning points
occur when
\begin{equation}
\label{eq16}
\sum_{D,E \subseteq \Xb}\frac{F_{D,E}}{S_{D,E}}\frac{\partial S_{D,E}}{\partial{x_i}}=0\hspace{1cm} (i=1,2,3).
\end{equation}
 In contrast to previous works~\citep{CHP01,CS04,CHHP,CKS03} that
operated in the space of the expected sequence variables, $S_{D,E}$,
here we are operating in the space of the path-set variables. This
eliminates the need to introduce the constraint of the ML points being
on a ``tree surface''.

From Equation (\ref{eq16a}) we see that many common $S_{D,E}$ values
are the same (e.g. $S_{1,\emptyset}=S_{1,1}=S_{\emptyset,1}$).
Substituting in (\ref{eq16}),  the turning points must satisfy:
\begin{equation}
\label{eq19}
\sum_{j=0}^4\frac{f_j}{a_j}\frac{\partial a_j}{\partial{x_i}}=0, \mbox{ for } i=1,2,3,
\end{equation}
where
\begin{eqnarray*}
f_0&=&F_{\0,\0},\\
f_1&=&F_{\0,1}+F_{1,\0}+F_{1,1},\\
f_2&=&F_{\0,2}+F_{2,\0}+F_{2,2},\\
f_3&=&F_{\0,12}+F_{12,\0}+F_{12,12},\\
f_4&=&F_{1,2}+F_{1,12}+F_{2,1}+F_{2,12}+F_{12,1}+F_{12,2}.\\
\end{eqnarray*}

We require our ML tree to adhere to the molecular clock
assumption, so a $((1,2),3)$-triplet tree under the molecular
clock assumption, must satisfy $q_1=q_2$ (see
Figure~\ref{fig:JC-triplet-with-q}).  We emphasise that the system
of equations does not take explicitly into account the {\em
inequality} $q_{3}\geq q_1$. The system is hard enough to solve as
it is. Of course, the final ML point (solution) should satisfy
this inequality, as otherwise it would not correspond to a
``real'' phylogenetic tree. In case no solution is found where
this inequality holds, the ML point should be sought on the
boundaries of the valid tree region.

\begin{figure}
\begin{center}
\input{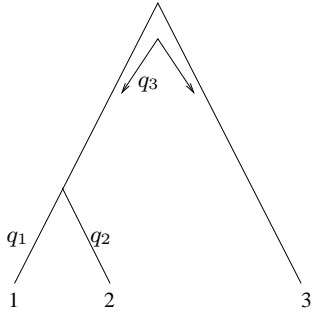}
\end{center}
\caption{A triplet tree under the molecular clock satisfies $q_1=q_2$.
\label{fig:JC-triplet-with-q}}
\end{figure}

From this
key identity, it follows that $x_1=x_2$ (see Equation (~\ref{eq:path-set}) for the
definition of the $x$ variables).
Therefore the likelihood
function (Equation~(\ref{eq15})) can be parameterised by the two free
variables $x_2$ and $x_3$,

and on the $f_i$ values derived from the observed site pattern
frequencies $F_{D,E}$. However under the constraint $x_1=x_2$ we can further
reduce the data by summing $f_1$ and $f_2$ so we define:
\begin{eqnarray*}
g_{00}&=&f_0=F_{\0,\0},\\
g_{01}&=&f_1+f_2=F_{\0,1}+F_{1,\0}+F_{1,1}+F_{\0,2}+F_{2,\0}+F_{2,2},\\
g_{03}&=&f_3=F_{\0,12}+F_{12,\0}+F_{12,12},\\
g_{04}&=&f_4=F_{1,2}+F_{1,12}+F_{2,1}+F_{2,12}+F_{12,1}+F_{12,2}.\\
\end{eqnarray*}

Consequently, Equations~(\ref{eq19})
reduce to  only two rational functions in the
two variables, $x_2$ and $x_3$ and the new parameters $g_{00}$,
$g_{01}$, $g_{03}$ and $g_{04}$.
In order to find the turning points we
take the numerators of these two equations. \\
} 

{Given the observed frequencies, $F_{D,E}$, of each site pattern $(D,E)
\subseteq \Xb \times \Xb$
(normalised so that $\sum_{D,E \subseteq \Xb}F_{D,E}=1$), then for any
expected sequence spectrum $S$ of some tree $T$, the likelihood of obtaining
those normalised frequencies is
\begin{equation}
\label{eq15}
L(F|T,S)=\prod_{D,E \subseteq \hat X}S_{D,E}^{F_{D,E}}.
\end{equation}
(It is convenient to use normalised frequencies as it simplifies the
formulae later. This normalisation
scales $\log L$ by a constant factor, so does not affect the identity of the
turning points.) Equation (10) gives identities among the pattern
probabilities $S_{D,E}$ so grouping the common factors in equation (12)
gives
\begin{equation}
\label{ }
L(F|T,S)=\prod_{j=0}^4a_j^{f_j},
\end{equation}
where
\begin{eqnarray*}
f_0&=&F_{\emptyset,\emptyset},\\
f_1&=&F_{\emptyset,1}+F_{1,\emptyset}+F_{1,1},\\
f_2&=&F_{\emptyset,2}+F_{2,\emptyset}+F_{2,2},\\
f_3&=&F_{\emptyset,12}+F_{12,\emptyset}+F_{12,12},\\
f_4&=&F_{1,2}+F_{1,12}+F_{2,1}+F_{2,12}+F_{12,1}+F_{12,2}.
\end{eqnarray*}
The expected sequence spectrum $S$ can be expressed as a function of the
three variables $x_1$, $x_2$ and $x_3$, so the values which maximise the
likelihood $L$ are obtained when the three partial derivatives,
$\frac{\partial L}{ \partial x_i}, (i= 1,2,3)$, are zero.
 In contrast to previous works~\citep{CHP01,CS04,CHHP,CKS03} that
operated in the space of the expected sequence variables, $S_{D,E}$,
here we are operating in the space of the path-set variables. This
eliminates the need to introduce the constraint of the ML points being
on a ``tree surface''.
By the chain rule, we get:
\begin{equation}
\label{eq:partial}
\frac{\partial L}{\partial x_i}=L \cdot \sum_{j=0}^4\frac{f_j}{a_j}
\frac{\partial a_j}{\partial x_i}=0, \mbox{ for } i=1,2,3.
\end{equation}

We require our ML tree to adhere to the molecular clock assumption, so
a $((1,2),3)-$triplet tree under this assumption requires $q_1=q_2 \le
q_3$ (see Figure~\ref{fig:JC-triplet-with-q}) which implies $x_1=x_2 \ge
x_3$.  In our analysis below we will explicitly impose the equality to
find the turning points. The inequality will need to be tested on any
potential solution, and if it were not satisfied, a maximum could be
sought on the boundary of the valid tree domain, where $x_1=x_2=x_3$.

\begin{figure}
\begin{center}
\input{JC-triplet-with-q.pstex_t}
\end{center}
\caption{A triplet tree under the molecular clock satisfies $q_1=q_2$.
\label{fig:JC-triplet-with-q}}
\end{figure}

The constraint $x_1=x_2$ implies $a_1=a_2$, so by setting
$f_{12}=f_1+f_2$ and $a_{12}=a_1=a_2$ we reduce the complexity of
equation (\ref{eq:partial}) to give two rational equations in two free
variables and the parameters $f_j$:
\begin{equation}
\label{ }
\frac{\partial L}{\partial x_i}=L \cdot
\left(\frac{f_0}{a_0}\frac{\partial a_0}{\partial x_i}+
\frac{f_{12}}{a_{12}}\frac{\partial a_{12}}{\partial x_i}+
\frac{f_3}{a_3}\frac{\partial a_3}{\partial x_i}+
\frac{f_4}{a_4}\frac{\partial a_4}{\partial x_i} \right)=0, \mbox{ for }
i=2,3.
\end{equation}

These simultaneously vanish when the two numerators, which are
polynomials in $x_2$, $x_3$ and the parameters $f_j$, are both
zero. We refer to these polynomial equations as $E_1$ and $E_2$.
} 

We now show that the  system of two resulting polynomials $\{E_1,E_2\}$ has only finitely many solutions, all of which
we can find.  The major tool used here is  the {\em resultant} of two
polynomials.
  Let $f(x) = \sum_{i=0}^d a_i x^i$ and $g(x) = \sum_{j=0}^d b_j x^j$
be two polynomials in one variable, $x$. The  resultant of $f$ and
$g$, denoted $Res(f,g, x)$, is a polynomial in the coefficients $a_i$
and $b_j$ of $f$ and $g$, which is 0 whenever $f$ and $g$ have a common
zero. The coefficients can themselves be unknowns, or
functions of other variables, in which case the resultant  replaces
the two polynomials $f$ and $g$ by a single polynomial in one fewer
variable.

Computing the resultant is a classical technique
for eliminating one variable from two equations. There is an elegant
formula for computing it due to Sylvester, and another due to Bezout, which have been
implemented in most computer algebra packages, such as {\tt Maple}.

We can compute the resultant $ER = Res(E_1, E_2, x_3)$ of $E_1$ and
$E_2$ with respect to $x_3$. This eliminates
$x_3$ from the equations and yields a single polynomial $ER$, in just
$x_2$ and the parameters. The polynomial $ER$ has the form:
\small{
\begin{eqnarray}
\label{eq:res}
\nonumber
ER&=&k\,{\it f_{3}}\,{\it f_{12}}\,{\it f_{0}}\,x_2^{13}{\it f_{4}}\,
 \left( 3\,x_{{2}}+1 \right)  \left( 2\,x_{2}^{2}+x_2+1
 \right)  \left( 3\,x_{{2}}^{2}+1 \right)  \left( 3\,x_2^{2}+3
\,x_{{2}}+2 \right) \\
&& \left( x_{{2}}-1 \right) ^{2} \left( x_{{2}}+1
 \right) ^{3}\cdot P_0
\end{eqnarray}
}
where $P_0$ is a degree $11$ polynomial with $288$ monomials and $k$
is some big constant.
\begin{theorem}
The turning points of $L$ (equation~\ref{eq15}) corresponding to
realistic  trees (namely, trees with positive edge lengths) are exactly the roots of
$P_0$.
\end{theorem}
\begin{proof}
The only term in $ER$ except for $P_0$ (Equation~\ref{eq:res}) that
admits positive real roots is the term $\left(
  x_{{2}}-1 \right)$. However, by the definition of $x_2$, this
root corresponds to $q_2=0$ which is not a realistic tree.
\end{proof}

\begin{corollary}
The Jukes-Cantor triplet has a finite number of ML points.
\end{corollary}
\begin{proof}
$P_0$ has at most $11$ different solutions and for each such a
solution we back substitute to obtain all the values of $x_3$.
\end{proof}

\section{Results on Genomic Sequences}
\label{sec:experiment} In order to evaluate our method, we tested
it on real genomic sequences. We looked at the NK cell receptor D
gene on human, mouse and rat (accession numbers AF260135, AF030313
and AF009511 respectively). We aligned the sequences using
CLUSTALW \citep{NAR:ThompsonHG1994}. Next, we computed the observed
sequence spectrum, as explained in Section~\ref{sec:Hadamard} and
illustrated in Table~\ref{tab:table1}. Three sequences have $16$
site patterns and therefore the observed sequence spectrum is
written in a $4$-by-$4$ matrix. The resulting spectrum is shown in
Table~\ref{tab:observed}.

\begin{table}[h]
\begin{center}
{\small
\begin{tabular}{|c||c|c|c|c|}
\hline
pattern &&&&\\
frequency & 00 & 01 & 10 & 11 \\
\hline
\hline
00 & $424$ & $18$ & $18$ & $80$ \\
\hline
01 & $1$ & $7$ & $2$ & $2$\\
\hline
10 & $7$ & $4$ & $4$ & $4$\\
\hline
11 & $27$ & $1$ & $2$ &$40$\\
\hline
\end{tabular}
} 

\end{center}
\caption{The observed sequence spectrum of NK cell receptor D gene of human, mouse and rat}
\label{tab:observed}
\end{table}

We calculated the maximum likelihood value for each of the three
rooted trees under the model for the three species. As expected the
((rat,mouse),human) tree was maximal, with edge lengths
$q_1=q_2=0.0197$ to rat and mouse and $q_3=0.1061$ to human, giving the
log likelihood $\ln L = -870.2$.

\remove{
For three species there are three different rooted trees, and we tested all three.
As is to be expected, the highest likelihood triplet is
the ((rat,mouse),human)  tree. Likelihood is maximized for edge lengths $q_1=q_2=0.0197$ and $q_3=0.1061$,  with log likelihood $-870.2$.
}

We also calculated the maximum likelihood value for each of the three
rooted trees for the beta actin gene for the three species guinea pig,  goose
and C elegans,(acc. numbers AF508792, M26111 and NM\_076440
resp.) finding the ((guinea pig, goose), C
elegans) tree maximal, with $q_1=q_2=0.021819$ and $q_3=0.050188$ giving $\ln
L = -1241.5$. Finally we calculated the maximum likelihood value for
each of the three rooted trees for the histone gene of Drosophila
melangoster, Hydra vulgaris and Human (acc numbers AY383571,
AY383572 and NM\_002107 resp.) finding the
((D. melangoster, H. vulgaris),Human) tree maximal, with
$q_1=q_2=0.001555$ and $q_3=0.012740$ with $\ln L = -86.835133$.

Each of the results above agree closely with the numerical values
obtained using the popular phylogenetic reconstruction packages
Phylip~\citep{CLAD:Felsenstein1989} and PAUP*~\citep{pau} which use
iterative methods to estimate the maxima.

\remove{
We also tested for the beta actin gene for the species guinea pig,
goose and  C elegans (acc. numbers AF508792, M26111 and NM\_076440
resp.). Using our formulae, we
obtained the same tree with slightly different edge lengths
$q_1=q_2=0.021819$ and $ q_3=.050188$ with log likelihood $-1241.5$.
The last test we conducted was on the histone gene for the Drosophila
melanogaster, Hydra vulgaris and human (acc. numbers AY383571,
AY383572 and NM\_002107 resp.). The tree obtained  was
the $((Drosophila, Human), Hydra)$ with edge lengths
 of $q_1= q_2=0.001555$ and
$q_3=.012740 $ and log likelihood $-86.835133$.

These results agree with those obtained by the popular phylogenetic
reconstruction packages,
Phylip~\citep{CLAD:Felsenstein1989} and PAUP*~\citep{pau}.
}

\section{Directions for Future Research}
\label{sec:conc}
The progress made here brings up a number of open problems:
\begin{itemize}
\item Our ML solutions are derived from the roots of a univariate, degree 11 polynomial.
This implies that the number of ML solutions is finite. It would
be interesting to explore the question of {\em uniqueness} of the
solution. If this is the case, it will most likely follow from the
existence of a single solution corresponding to a realistic tree,
as in~\citep{CKS03}.
\item The Jukes-Cantor substitution model is the a special case of the family of Kimura
substitution models. It would be interesting to further extend the
result in this paper for the other models (two and three parameters)
of the Kimura family.
\item It would be interesting to extend these results to rooted trees with {\em four leaves} under JC model and a molecular clock. Here we have two different topologies -- the fork and the comb~\citep{CKS03}.
It is expected that such extension will face substantial  technical difficulties.
\end{itemize}

\bigskip

\noindent {\bf Acknowledgements:} Thanks to Joseph Felsenstein for
fruitful discussions and to Bernd Sturmfels for
enlightening comments on this manuscript and informing us about
~\citep{HKS04}.

\bibliographystyle{elsart-harv}
\bibliography{fork}

\clearpage
\end{document}